\documentclass{article}
\usepackage{amsmath,amssymb,graphicx,subfig,epsfig,psfrag,color,mathrsfs}

\newcommand{\siun}[1]{\mathrm{#1}} 
\newcommand{\lambert}{\mathscr{W}}
\newcommand{\arcsinh}{\mathop{\rm arcsinh}}
\newcommand{\scri}{\mathscr{I}}
\newcommand{\CITE}[1]{\cite{#1}}

\newcommand{\re}[1]{(\ref{eq:#1})}

\def\phi{\varphi}

\def\d{\mathrm{d}}
\def\p{\partial}
\def\cosh{\mathop{\rm cosh}}
\def\sinh{\mathop{\rm sinh}}

\def\diag{\mathop{\rm diag}}

\begin{document}

\title{The Newtonian limit of spacetimes for accelerated particles and black holes}
\author{Ji\v{r}\'i Bi\v{c}\'ak \and David Kofro\v{n}}

\maketitle

\begin{abstract}
Solutions of vacuum Einstein's field equations describing uniformly accelerated particles or black holes  belong to the class of boost-rotation symmetric spacetimes. They are the only explicit solutions known which represent moving finite objects. Their Newtonian limit is analyzed using the Ehlers frame theory. Generic spacetimes with axial and boost symmetries are first studied from the Newtonian perspective. The results are then illustrated by specific examples such as C-metric, Bonnor-Swaminarayan solutions, self-accelerating ``dipole particles'', and generalized boost-rotation symmetric solutions describing freely falling particles in an external field. In contrast to some previous discussions, our results are physically plausible in the sense that the Newtonian limit corresponds to the fields of classical point masses accelerated uniformly in classical mechanics. This corroborates the physical significance of the boost-rotation symmetric spacetimes.
\end{abstract}

\section{Introduction}
There is only one class of explicitly known exact solutions of vacuum Einstein's field equations which represent moving finite sources: it describes ``uniformly accelerated particles or black holes''. In a curved spacetime the ``uniform acceleration'' is understood with respect to a fictitious flat background. However, these solutions can be characterized fully geometrically since, in addition to the Killing vector associated with axial symmetry, they admit a second Killing vector which becomes a boost in the flat-space limit. Therefore, in \CITE{Bicak-TGS} the name ``boost-rotation symmetric spacetimes'' was coined, and this has beeen used in the literature thereafter -- among others in the second edition of the ``Exact -- Solutions book'' \CITE{SC}, where some of the solutions with boost-rotation symmetry are briefly described in Section 17.2.

The first solutions of this type were obtained by Bonnor and Swaminayaran \CITE{BSzp} and Israel and Khan \CITE{nevim2} in 1964. In 1968 Bi\v{c}\'ak \CITE{Bicak68} shown that these solutions are radiative. Together with other boost-rotation symmetric solutions discovered later, they are the only explicit examples of spacetimes  with gravitational radiation described by a non-vanishing Bondi's news function, they exhibit peeling-off property and admit asymptotically flat null infinity $\scri$ satisfying Penrose's requirements, at least locally \CITE{1984JMP....25..600B}, \CITE{nevim3}. In fact, this situation may not change soon since it appears extremely complicated to search for such explicit radiative solutions with only one symmetry, and among those with two symmetries, the boost-rotation symmetry plays a unique role: Theorems can be proven that in axially symmetric, locally asymptotically flat spacetimes (so that $\scri$ exists though not necessarily  globally), the only \emph{additional} symmetry that does not exclude radiation is the \emph{boost} symmetry\footnote{See \CITE{1984JMP....25..600B} for the precise formulation and the proof of the theorem, and \CITE{nevim3} for the generalization to the electrovacuum spacetimes with Killing vectors which need not be hypersurface orthogonal.}.

In Minkowski spacetime the boost Killing vector associated with the boost along the $z$-axis has the form
\begin{equation}
\zeta_{boost} = z\,\frac{\p}{c\p t}+ct\,\frac{\p}{\p z}\,.
\label{eq:boostKV}
\end{equation}
Here we have to keep the speed of light $c$ because of our interest in the Newtonian limit. The orbits of symmetry to which the Killing vector is tangent are hyperbolas $z^2-c^2t^2=B=$~constant, $x,\,y=$~constant. Orbits with $B>0$ are timelike. They can represent worldlines of uniformly accelerated particles in special relativity. If a point charged (electric or scalar) particle moves along the $z$-axis with a uniform acceleration and is momentarily at $t=0$ at rest at $z=\sqrt{B}$, its worldline is the hyperbola $x=y=0$, $z=\sqrt{c^2t^2+B}$, $c^2/\sqrt{B}$ is its acceleration; its (scalar or electromagnetic) field has boost-rotation symmetry. The field is analytic everywhere outside the source if there are \emph{two} particles located symmetrically with respect to the plane $z=0$ and accelerated in opposite directions, and both retarded and advanced effects are admitted.

\begin{figure}
\begin{center}
\subfloat{
%
%
\begin{psfrags}%
\psfragscanon%
%
\psfrag{s05}[l][l]{\color[rgb]{0,0,0}\setlength{\tabcolsep}{0pt}\begin{tabular}{l}$z$\end{tabular}}%
\psfrag{s06}[l][l]{\color[rgb]{0,0,0}\setlength{\tabcolsep}{0pt}\begin{tabular}{l}$r$\end{tabular}}%
%
\psfrag{x01}[t][t]{-1}%
\psfrag{x02}[t][t]{-0.5}%
\psfrag{x03}[t][t]{0}%
\psfrag{x04}[t][t]{0.5}%
\psfrag{x05}[t][t]{1}%
%
\psfrag{v01}[r][r]{-1}%
\psfrag{v02}[r][r]{-0.5}%
\psfrag{v03}[r][r]{0}%
\psfrag{v04}[r][r]{0.5}%
\psfrag{v05}[r][r]{1}%
%
\includegraphics[keepaspectratio,height=6cm]{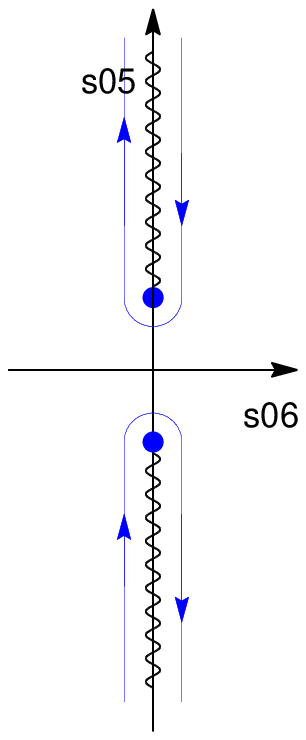}%
\end{psfrags}%
%
}\qquad\qquad
\subfloat{
%
%
\begin{psfrags}%
\psfragscanon%
%
\psfrag{s05}[l][l]{\color[rgb]{0,0,0}\setlength{\tabcolsep}{0pt}\begin{tabular}{l}$t$\end{tabular}}%
\psfrag{s06}[l][l]{\color[rgb]{0,0,0}\setlength{\tabcolsep}{0pt}\begin{tabular}{l}$z$\end{tabular}}%
%
\psfrag{x01}[t][t]{-1}%
\psfrag{x02}[t][t]{-0.5}%
\psfrag{x03}[t][t]{0}%
\psfrag{x04}[t][t]{0.5}%
\psfrag{x05}[t][t]{1}%
%
\psfrag{v01}[r][r]{-1}%
\psfrag{v02}[r][r]{-0.5}%
\psfrag{v03}[r][r]{0}%
\psfrag{v04}[r][r]{0.5}%
\psfrag{v05}[r][r]{1}%
%
\includegraphics[keepaspectratio,height=6cm]{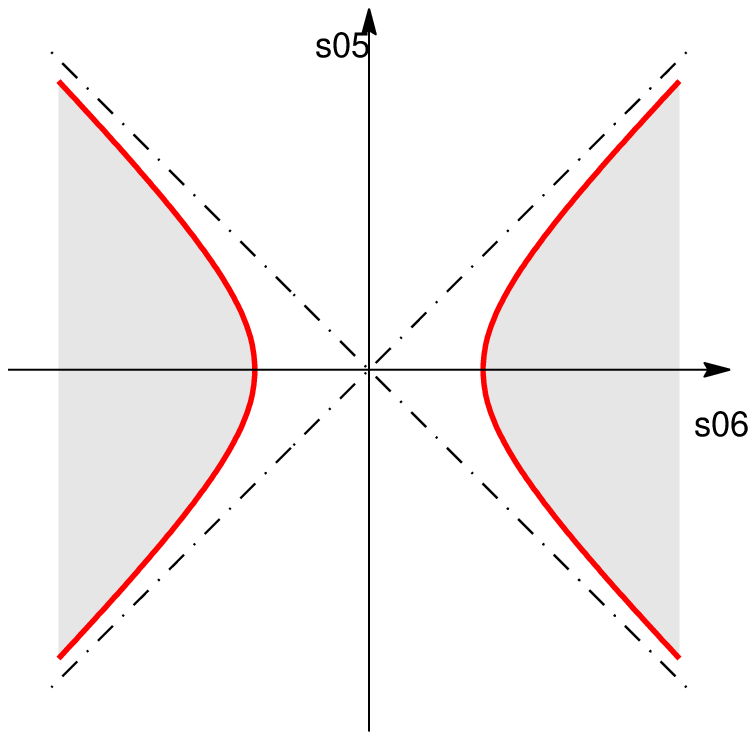}%
\end{psfrags}%
%
}
\end{center}
\caption{Space (a) and spacetime (b) diagram with two uniformly accelerated particles. The ``strings'' are depicted in (a) by zigzag lines, their history is in the shaded region (b).}
\label{fig:1}
\end{figure}

Such particles are shown in space and spacetime diagram in Fig. \ref{fig:1}. From the figure it is evident that the two particles move independently of each other since their worldlines are separated by the null hypersurfaces $z=\pm \, ct$. Following \CITE{Bicak-TGS}, we call the null hypersurfaces $z=\pm \, ct$ the ``roof''. The boost Killing vector becomes null on these hypersurfaces, it is timelike for $c^2t^2<z^2$, i.e., ``below the roof'', but it is spacelike ``above the roof'' ($c^2t^2>z^2$) -- cf. Fig. \ref{fig:1}.  Since the particles' worldlines approach null lines both in the past and future, it is intuitively clear that they ``start'' at the past null infinity $\scri^-$ and ``end up'' at the future null infinity $\scri^+$.

These features of the boost-rotation symmetric fields in Minkowski space emerge without a change also in the curved spacetimes with boost-rotation symmetry in general relativity. From a unified point of view these spacetimes were defined and treated geometrically in \CITE{Bicak-TGS} (see also reviews \CITE{bicak-ssefe}, \CITE{2000CzJPh..50..333P}). We refer to this detailed work for rigorous definitions and theorems, including the global analysis and investigation of the properties of null infinity $\scri$. In general relativity, the ``causes'' of the motion are incorporated in the theory. In a general case of the boost-rotation symmetric solutions there occur conical singularities of the metric distributed along the $z$-axis (see Fig. \ref{fig:1}). They can be considered as ``strings'' (or ``rods'') which cause the particles to accelerate. They appear also at $\scri$, so that some parts of its generators are missing; however, the distribution of nodes/strings can always be arranged in such a way that $\scri$ admits smooth regular sections \CITE{Bicak-TGS}. In the exceptional cases, when $\scri$ is regular except for four points, either the particles are  ``self-accelerating'' due to their ``inner'' structure which has to include a negative mass, or there are more particles distributed along $z>0$, and symmetrically along $z<0$, with the signs and the magnitudes of their masses and accelerations properly chosen. An infinite number of solutions with self-accelerating particles was constructed in \CITE{1983RSPSA.390..411B}; a pair of chasing particles with positive and negative mass was considered approximately in the region $z^2>c^2t^2$ first by Bondi \CITE{bondi-nm} with the particles being ``real extended bodies''; the corresponding complete exact solution with particles represented by (Curzon-type) singularities, is included in \CITE{BSzp} and its radiative properties are analyzed in \CITE{Bicak68}. 

In the exact boost-rotation symmetric solutions available explicitly the ``particles'' or ``sources'' can be represented not only by singularities but also by black holes. One can construct two or arbitrary even number of oppositely located uniformly accelerated black holes connected by a string which pulls them apart. Then $\scri$ admits regular sections and spatial infinity is regular. Alternatively, each black hole is attached to a cosmic string extending to infinity, the region ``between the holes'' is regular. In general, conical singularities extend along the whole $z$-axis but that between the holes has a smaller deficit angle;  when the conical singularities are interpreted as cosmic strings, the difference in string tensions provides the cause of the acceleration. All these cases are described by the well-known C-metric; see, e.g. \CITE{PhysRevD.2.1359}, \CITE{bonnorscm}, \CITE{2003CQGra..20..127D}, \CITE{0264-9381-23-2-005}, and references therein. The distribution of the strings and the sources along the $z$-axis can be interpreted for the particles represented by singularities and black holes in the same way. There also exist ``generalized'' boost-rotation symmetric spacetimes which contain neither cosmic strings nor negative masses. They describe accelerated particles in asymptotically ``uniform'' external fields which can be obtained by appropriate limiting procedures from asymptotically flat solutions \CITE{1983RSPSA.390..397B}, \CITE{1983RSPSA.390..411B}. All the special cases of the boost-rotation symmetric spacetimes mentioned above will be considered in the following. 

Although much understanding of the spacetimes with boost-rotation symmetry has been gained, some possible roles of these solutions in general relativity have not yet been fully elucidated. Various specific boost-rotation symmetric solutions were employed as test beds in numerical relativity based on a null-cone formalism \CITE{BRW}, \CITE{GPW}, and on a standard spacelike initial hypersurface \CITE{AGS}. Already in \CITE{Bicak68}, where the radiative character of the specific boost-rotations symmetric solutions was first discovered and in a ``weak-field limit'' compared with radiation from an analogous electromagnetic system, it was suggested that these solutions could be used as tests of analytical approximation methods. The first natural step in a systematic comparison with post-Newtonian approximation methods is to investigate the Newtonian limit.

Numerous useful, though frequently heuristic, studies of such a limit appeared since the birth of general relativity. The most geometric -- and from various viewpoints most profound -- treatment was initiated independently by Cartan \CITE{Cartan1}\,--\CITE{Cartan2} and Friedrichs \CITE{Fr} in 1920's with their formulation of Newton's theory in a generally covariant 4-dimensional spacetime language. This so-called Newton-Cartan theory led to the Ehlers frame theory \CITE{EhlersN}, \CITE{0264-9381-14-1A-010}, \CITE{E2}. The frame theory is encompassing general relativity and Newton-Cartan theory. Our treatment is based on this theory.

The Newtonian limit of the boost-rotation symmetric spacetimes employing the Ehlers frame theory was, in fact, recently investigated by Lazkoz and Valiente Kroon \CITE{lazkoz-2004-460}. They review the Ehlers frame theory in detail, apply it carefully to the general boost-rotation symmetric spacetimes, and illustrate by specific examples. However, the results they obtain exhibit surprisingly ``unphysical features'', using their own words: ``The interpretation of particles moving in a uniformly accelerated  fashion is only valid for early times $t\approx 0$.'' Or elsewhere: ``Thus the Newtonian limit for early times is a strictly static Newtonian potential in which the sources are not moving.''
Moreover ``the boost Killing vector field is not inherited'' in the Newtonian limit, although one would expect that it will just go over to the boosts in the Galilei group. The authors of \CITE{lazkoz-2004-460} attribute these unphysical features of the Newtonian limits to ``problems already existing in the general relativistic solutions''.

In the present work we show that one can construct the Newtonian limits with entirely plausible properties. How can such seemingly conflicting conclusions be understood without finding a mathematical inconsistency in one of the treatments? The answer lies in the fact that \emph{the construction of a Newtonian limit is not a uniquely prescribed procedure}. First, one distributes ``appropriately''  parameter $\lambda=c^{-2}$ into both the starting general-relativistic metric and matter variables. In this way an one-parameter family of spacetimes is constructed. The Newtonian limit is then obtained by letting $\lambda\rightarrow 0$. However, there exists no \emph{a priori} prescription for the construction of this family (or, the ``allocation'' of $\lambda$'s at the beginning). In such a situation the best guidance is to anticipate a ``reasonable'' Newtonian limit and to construct the family of spacetimes accordingly. 

Now Lazkoz and Kroon \CITE{lazkoz-2004-460} motivated their choice as follows: 

\begin{footnotesize}
``The natural arena for the discussion of the Newtonian limit of boost-rotation symmetric spacetimes is the region above the roof, where the spacetime is radiative. The reason for this is that as one makes $\lambda\rightarrow 0$, the region below the roof gets squeezed by the roof. That is, the region below the roof disappears in the limit, while the roof becomes the set $\{t=0\}$.''
\end{footnotesize}

Here, it is, of course, correctly pointed out that the null hypersurfaces of the roof degenerate into a spacelike surface in the limit $\lambda\rightarrow 0$ (like null cones get flattened with $c\rightarrow\infty$). However, to stay in the radiative part of the boost-rotation symmetric spacetimes means to ``loose the particles''! No wonder that the resulting limit has unphysical features. 

In our approach we stay \emph{below} the roof, i.e., in the region where the boost Killing vector is timelike and where the sources occur. In order not to get ``squeezed'' by the roof as $\lambda\rightarrow 0$, we have to ``run away'' with the sources. In accordance with this idea we parametrize appropriately the particles' worldlines. In this way we arrive at the Newtonian limit with all the features expectable within the Newtonian gravity. 

In Section~\ref{sec:theory} we first very briefly summarize the Newtonian limit as it is formulated in Ehlers's frame theory \CITE{E2}\,--\,\CITE{EhlersN}. Since the detailed expositions of this theory are not easily accessible, the reader may wish to consult the work of Lazkoz and Valiente Kroon \CITE{lazkoz-2004-460} where the frame theory is well described, including rigorous definitions, theorems and references to other works related to the theory. We then explain in detail our approach to the Newtonian limit of general boost-rotation symmetric spacetimes. The Newtonian limits of several typical examples of the solutions with uniformly accelerated sources, in particular those mentioned above, are studied in Section~\ref{sec:examples}. Some concluding remarks comprise Section~\ref{sec:conclusions}. We also add an Appendix in which the Newtonian limit of the Schwarzschild solution is studied within the whole Schwarzschild-Kruskal manifold. There are well-known similarities between the Schwarzschild horizon and the acceleration horizon (the roof), so that the comparison of the Newtonian limits is illuminating.

\section{The Newtonian limit of the boost-rotation symmetric spacetimes: The general case}\label{sec:theory}
In contrast to \CITE{lazkoz-2004-460}, our construction of the Newtonian limit does not start from the set of observers which fill in the region ``above the roof'' $(c^2t^2>z^2)$ where no timelike Killing vector exists and no sources occur. Rather, we consider all timelike worldlines with fixed canonical coordinates $\{x,\,y,\,z\}$ (see \CITE{Bicak-TGS} for their geometrical meaning), i.e., equivalently, with fixed $\{\rho,\,\phi,\,z\}$ related to them as cylindrical to cartesian coordinates in flat space. Our starting form of the metric in the canonical coordinates is
\begin{equation}
\d s^2 = \frac{ e^{\mu}c^2\left(z\,\d t-t\,\d z\right)^2-e^{\nu} \left(z\,\d z-c^2t\,\d t\right)^2}{z^2-c^2t^2}\ - e^{2\nu}\,\d \rho^2-e^{-\mu}\rho^2\,\d \phi^2 \,,
\label{eq:metric}
\end{equation}
where $\mu=\mu(\rho^2,\,c^2t^2-z^2)$, $\nu=\nu(\rho^2,\,c^2t^2-z^2)$ and the coordinate ranges are $t,\,z\in(-\infty,\,\infty)$, $\rho\in[0,\,\infty)$, $\phi\in[0,\,2\pi)$.

This metric is global -- it is valid both above and below the roof (not just above it as stated in \CITE{lazkoz-2004-460}). As a consequence of vacuum Einstein's equations, function $\mu$ satisfies the ordinary flat-space wave equation 
\begin{equation}
\square \mu = \left[ -\frac{1}{c^2}\frac{\p^2}{\p t^2} + \frac{1}{\rho}\frac{\p}{\p\rho}\left(\rho\frac{\p}{\p\rho}\right) + \frac{\p^2}{\p z^2} \right] \mu = 0\,.
\label{eq:wave-eq}
\end{equation}
The other metric function\footnote{In \CITE{Bicak-TGS} and other references this function is usually called $\lambda$. We use $\nu$ instead of $\lambda$ because $\lambda$ is here reserved for the causality constant of Ehlers's frame theory.}  $\nu$ is determined by nonlinear equations 
\begin{eqnarray}
\left( A+B \right)\nu_{,A} &=& B\left( A\mu^2_{,A}-B\mu^2_{,B}+2A\mu_{,A}\,\mu_{,B} \right)\nonumber\\
&&\qquad +\left( A-B \right)\mu_{,A}-2B\mu_{,B} \,, \label{eq:nuA} \\
\rule[.2em]{0pt}{1em}
\left( A+B \right)\nu_{,B} &=& A\left( B\mu^2_{,B}-A\mu^2_{,A}+2B\mu_{,A}\,\mu_{,B} \right)\nonumber\\
&&\qquad +\left( A-B \right)\mu_{,B}+2A\mu_{,A} \,, \label{eq:nuB}
\label{eq:nu-eq}
\end{eqnarray}
where $A=\rho^2,\, B=z^2-c^2t^2$. The integrability condition of Eqs. \re{nuA} and \re{nuB} is just the wave equation \re{wave-eq} which, written also in terms of $A$ and $B$, reads
\begin{equation}
A\mu_{,AA}+B\mu_{,BB}+\mu_{,A}+\mu_{,B}=0\,.
\label{eq:dAlembert}
\end{equation}

Following Ehlers's frame theory, we consider a family of 4-dimensional manifolds $\mathscr{M}(\lambda)$ parametrized by the ``causality constant'' $\lambda\,(=1/c^2 \text{ for } \lambda\neq 0)$. Each manifold is endowed with a torsion-free connection $\Gamma^\alpha_{\beta\gamma}$ (``gravitational field'') and with temporal and spatial metrics which are given by  symmetric tensors $t_{\alpha\beta}$ and $s^{\alpha\beta}$. These metrics satisfy the relation
\begin{equation}
t_{\alpha\sigma}s^{\sigma\beta} = -\lambda \delta^\beta_\alpha\,,
\label{eq:tsm}
\end{equation}
and are compatible with the connection $\Gamma^\alpha_{\beta\gamma}$. The frame theory requires that for $\lambda\neq 0$ the metrics are $t_{\alpha\beta} = -\lambda g_{\alpha\beta}$ and $s^{\alpha\beta} = g^{\alpha\beta}$.

In the limit $\lambda\rightarrow 0$ the frame theory goes over to the Newton-Cartan theory -- a geometrical formulation of the classical Newton theory (see, e.g., \CITE{Trautmann} and references quoted in \CITE{EhlersN}). Then there exists (at least locally) a Cartesian coordinate system such that the metrics have simple forms 
$$t_{\alpha\beta} = \diag (1,\,0,\,0,\,0)\,, \qquad s^{\alpha\beta} = \diag (0,\,1,\,1,\,1)\,,$$
and the gravitational field can be expressed by means of the affine connection as
\begin{equation}
\Gamma^\alpha_{\beta\gamma} = t_{,\beta}\,t_{,\gamma}\,s^{\alpha\delta}\,\Phi_{,\delta}\,,
\label{eq:Aff}
\end{equation}
where $t$ is the global (scalar) absolute time and $\Phi$ is the Newtonian gravitational potential. 

We shall now derive the relevant quantities entering the frame theory for the boost-rotation symmetric spacetimes. First let us write the worldline of a uniformly accelerated particle in special relativity (see, e.g., \CITE{Rindler} for a pedagogical exposition) in the form
\begin{equation}
z = \sqrt{\frac{c^4}{g^2}+c^2t^2} = \sqrt{\frac{1}{\lambda^2g^2}+\frac{t^2}{\lambda}}\,.
\label{eq:uam}
\end{equation}
This describes the well-known hyperbolic motion when the particle starts at infinity with velocity approaching $c$, arrives at $z|_{t=0}=c^2/g=1/\lambda g$ where it stops, and then continues back to infinity, approaching it asymptotically with $v\rightarrow c$; $g$ is its proper acceleration.

\begin{figure}[!h]
\subfloat[$\lambda=1$]{
%
%
\begin{psfrags}%
\psfragscanon%
%
\psfrag{s05}[l][l]{\color[rgb]{0,0,0}\setlength{\tabcolsep}{0pt}\begin{tabular}{l}$t$\end{tabular}}%
\psfrag{s06}[l][l]{\color[rgb]{0,0,0}\setlength{\tabcolsep}{0pt}\begin{tabular}{l}$z$\end{tabular}}%
%
\psfrag{x01}[t][t]{-5}%
\psfrag{x02}[t][t]{0}%
\psfrag{x03}[t][t]{5}%
%
\psfrag{v01}[r][r]{-5}%
\psfrag{v02}[r][r]{0}%
\psfrag{v03}[r][r]{5}%
%
\includegraphics[keepaspectratio,width=2.6cm]{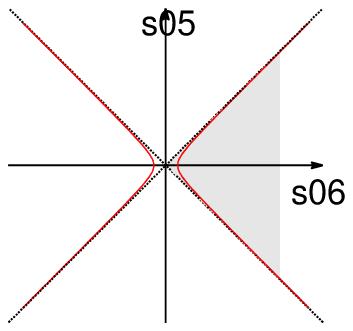}%
\end{psfrags}%
%
}\hfill
\subfloat[$\lambda=1/2$]{
%
%
\begin{psfrags}%
\psfragscanon%
%
\psfrag{s05}[l][l]{\color[rgb]{0,0,0}\setlength{\tabcolsep}{0pt}\begin{tabular}{l}$t$\end{tabular}}%
\psfrag{s06}[l][l]{\color[rgb]{0,0,0}\setlength{\tabcolsep}{0pt}\begin{tabular}{l}$z$\end{tabular}}%
%
\psfrag{x01}[t][t]{-5}%
\psfrag{x02}[t][t]{0}%
\psfrag{x03}[t][t]{5}%
%
\psfrag{v01}[r][r]{-5}%
\psfrag{v02}[r][r]{0}%
\psfrag{v03}[r][r]{5}%
%
\includegraphics[keepaspectratio,width=2.6cm]{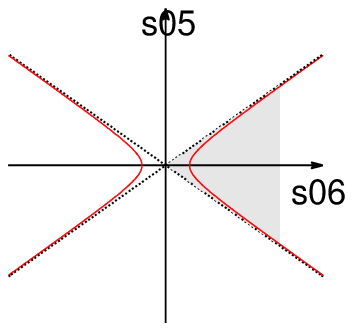}%
\end{psfrags}%
%
}\hfill
\subfloat[$\lambda=1/4$]{
%
%
\begin{psfrags}%
\psfragscanon%
%
\psfrag{s05}[l][l]{\color[rgb]{0,0,0}\setlength{\tabcolsep}{0pt}\begin{tabular}{l}$t$\end{tabular}}%
\psfrag{s06}[l][l]{\color[rgb]{0,0,0}\setlength{\tabcolsep}{0pt}\begin{tabular}{l}$z$\end{tabular}}%
%
\psfrag{x01}[t][t]{-5}%
\psfrag{x02}[t][t]{0}%
\psfrag{x03}[t][t]{5}%
%
\psfrag{v01}[r][r]{-5}%
\psfrag{v02}[r][r]{0}%
\psfrag{v03}[r][r]{5}%
%
\includegraphics[keepaspectratio,width=2.6cm]{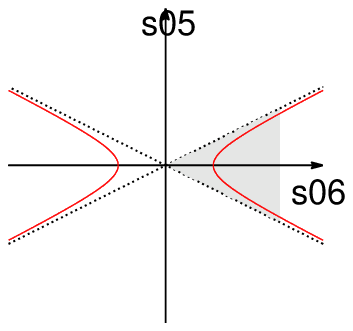}%
\end{psfrags}%
%
}\hfill
\subfloat[$\lambda=1/20$]{
%
%
\begin{psfrags}%
\psfragscanon%
%
\psfrag{s05}[l][l]{\color[rgb]{0,0,0}\setlength{\tabcolsep}{0pt}\begin{tabular}{l}$t$\end{tabular}}%
\psfrag{s06}[l][l]{\color[rgb]{0,0,0}\setlength{\tabcolsep}{0pt}\begin{tabular}{l}$z$\end{tabular}}%
%
\psfrag{x01}[t][t]{-5}%
\psfrag{x02}[t][t]{0}%
\psfrag{x03}[t][t]{5}%
%
\psfrag{v01}[r][r]{-5}%
\psfrag{v02}[r][r]{0}%
\psfrag{v03}[r][r]{5}%
%
\includegraphics[keepaspectratio,width=2.6cm]{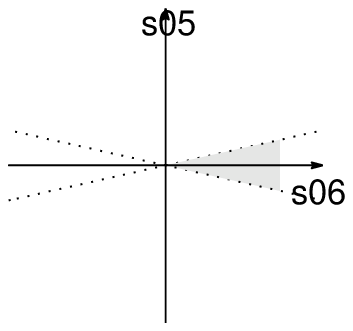}%
\end{psfrags}%
%
}
\\
\subfloat[$\lambda=1$]{
%
%
\begin{psfrags}%
\psfragscanon%
%
\psfrag{s05}[l][l]{\color[rgb]{0,0,0}\setlength{\tabcolsep}{0pt}\begin{tabular}{l}$t$\end{tabular}}%
\psfrag{s06}[l][l]{\color[rgb]{0,0,0}\setlength{\tabcolsep}{0pt}\begin{tabular}{l}$\zeta$\end{tabular}}%
%
\psfrag{x01}[t][t]{-5}%
\psfrag{x02}[t][t]{0}%
\psfrag{x03}[t][t]{5}%
%
\psfrag{v01}[r][r]{-5}%
\psfrag{v02}[r][r]{0}%
\psfrag{v03}[r][r]{5}%
%
\includegraphics[keepaspectratio,width=2.6cm]{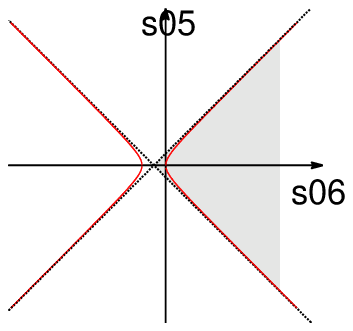}%
\end{psfrags}%
%
}\hfill
\subfloat[$\lambda=1/2$]{
%
%
\begin{psfrags}%
\psfragscanon%
%
\psfrag{s05}[l][l]{\color[rgb]{0,0,0}\setlength{\tabcolsep}{0pt}\begin{tabular}{l}$t$\end{tabular}}%
\psfrag{s06}[l][l]{\color[rgb]{0,0,0}\setlength{\tabcolsep}{0pt}\begin{tabular}{l}$\zeta$\end{tabular}}%
%
\psfrag{x01}[t][t]{-5}%
\psfrag{x02}[t][t]{0}%
\psfrag{x03}[t][t]{5}%
%
\psfrag{v01}[r][r]{-5}%
\psfrag{v02}[r][r]{0}%
\psfrag{v03}[r][r]{5}%
%
\includegraphics[keepaspectratio,width=2.6cm]{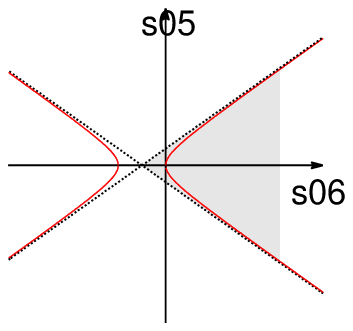}%
\end{psfrags}%
%
}\hfill
\subfloat[$\lambda=1/4$]{
%
%
\begin{psfrags}%
\psfragscanon%
%
\psfrag{s05}[l][l]{\color[rgb]{0,0,0}\setlength{\tabcolsep}{0pt}\begin{tabular}{l}$t$\end{tabular}}%
\psfrag{s06}[l][l]{\color[rgb]{0,0,0}\setlength{\tabcolsep}{0pt}\begin{tabular}{l}$\zeta$\end{tabular}}%
%
\psfrag{x01}[t][t]{-5}%
\psfrag{x02}[t][t]{0}%
\psfrag{x03}[t][t]{5}%
%
\psfrag{v01}[r][r]{-5}%
\psfrag{v02}[r][r]{0}%
\psfrag{v03}[r][r]{5}%
%
\includegraphics[keepaspectratio,width=2.6cm]{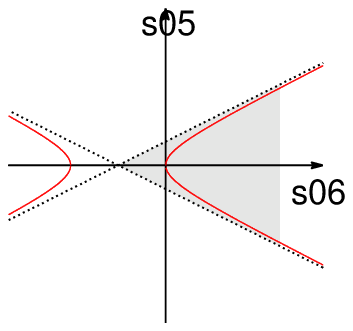}%
\end{psfrags}%
%
}\hfill
\subfloat[$\lambda=1/20$]{
%
%
\begin{psfrags}%
\psfragscanon%
%
\psfrag{s05}[l][l]{\color[rgb]{0,0,0}\setlength{\tabcolsep}{0pt}\begin{tabular}{l}$t$\end{tabular}}%
\psfrag{s06}[l][l]{\color[rgb]{0,0,0}\setlength{\tabcolsep}{0pt}\begin{tabular}{l}$\zeta$\end{tabular}}%
%
\psfrag{x01}[t][t]{-5}%
\psfrag{x02}[t][t]{0}%
\psfrag{x03}[t][t]{5}%
%
\psfrag{v01}[r][r]{-5}%
\psfrag{v02}[r][r]{0}%
\psfrag{v03}[r][r]{5}%
%
\includegraphics[keepaspectratio,width=2.6cm]{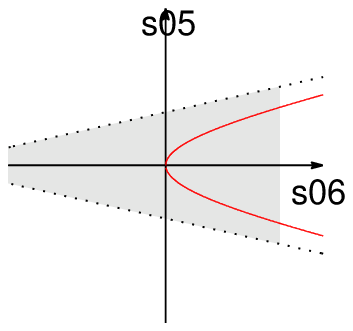}%
\end{psfrags}%
%
}
\caption{The roof and the particles worldlines for different values of $\lambda$ in the global coordinates $t-z$ (figures (a)\,--\,(d)) and for the same values of $\lambda$ in the shifted coordinates $t-\zeta$, where $\zeta=z-1/\lambda g$ (figures (e)\,--\,(h)). The corresponding regions are shaded.} 
\label{fig:roof}
\end{figure}

Applying the limit $\lambda\rightarrow 0$, we immediately observe that as $\lambda\rightarrow 0$ the null cone -- and so the roof -- becomes the hyperplane $t=0$, and the particle, including its turning point, is pushed away to spatial infinity. Hence, in order to obtain a nontrivial limit, we have to ``go'' to infinity with the particle. We do this by introducing a new coordinate $\zeta$ by 
\begin{equation}
z = \zeta + \frac{1}{\lambda g}\,,
\label{eq:subs}
\end{equation}
i.e., we make a $\lambda$-dependent shift of $z$. It is illustrated in Fig. \ref{fig:roof} how this shift works for various values of $\lambda$. Since in the boost-rotation symmetric spacetimes particles symmetrically located along $z<0$ occur, their worldlines are also plotted. However, notice that these are ``shifted away'' as $\lambda\rightarrow 0$.

We now apply this procedure to find the Newtonian limit of the boost-rotation symmetric spacetimes. Replacing the factor $c$ by $1/\sqrt{\lambda}$ in the metric \re{metric}, the metric tensor in coordinates $\{t,\,r,\,z,\,\phi \}$ turns out to be
\begin{equation}
g_{\alpha\beta} =
\begin{pmatrix}
\displaystyle\frac{1}{\lambda}\frac{e^\mu z^2 \lambda - e^\nu t^2}{z^2\lambda - t^2}  & 0 &\displaystyle -\frac{\left(e^\mu-e^\nu\right)tz}{z^2\lambda-t^2} & 0 \\
0 & -e^\nu & 0 & 0 \\
\displaystyle-\frac{\left(e^\mu-e^\nu\right)tz}{z^2\lambda-t^2} & 0 &  \displaystyle -\frac{e^\nu z^2\lambda -e^\mu t^2}{z^2\lambda-t^2}& 0 \\
0 & 0 & 0 &-r^2e^{-\mu}
\end{pmatrix} \,,
\label{eq:metric-mat} 
\end{equation}
its contravariant form is
\begin{equation}
g^{\alpha\beta} =
\begin{pmatrix}
\displaystyle \lambda\frac{e^{-\mu}z^2\lambda-e^{-\nu}t^2}{z^2\lambda-t^2} & 0 & \displaystyle \lambda \frac{tz\left( e^{-\mu}-e^{-\nu} \right)}{z^2\lambda-t^2} & 0 \\
0 & -e^{-\nu} & 0 & 0 \\
\displaystyle \lambda \frac{tz\left( e^{-\mu}-e^{-\nu} \right)}{z^2\lambda-t^2} & 0 & \displaystyle-\frac{e^{-\nu}z^2\lambda-e^{-\mu}t^2}{z^2\lambda-t^2} & 0 \\
0 & 0 & 0 & \displaystyle -\frac{e^{\mu}}{r^2} 
\end{pmatrix}\,,
\label{eq:invmetric-mat}
\end{equation}
Factor $\lambda^{-1}$ appears also in $\lambda^{-1}t^2$ in the arguments of functions $\mu$ and $\nu$. In this way the $\lambda$-factors are ``distributed'' uniquely in the metric and the Newtonian limit is thus fixed. Then, as noted above, the temporal and spatial metrics in the frame theory are simply $t_{\alpha\beta} = -\lambda g_{\alpha\beta}$ and $s^{\alpha\beta} = g^{\alpha\beta}$.

By analyzing all specific cases in Section~\ref{sec:examples} we shall see that when $\lambda\rightarrow 0$ the functions $\mu$ and $\nu$ are approaching constants and all their derivatives $\mu_{,\alpha}$ and $\nu_{,\alpha}$ vanish. This can be seen on physical grounds by regarding Eq.~\re{wave-eq} with a source\footnote{One of the Ricci tensor components turns out to be $R_{\phi\phi}=-\frac{1}{2}\,e^{-\mu-\nu}\rho^2\,\square\mu$, and the right-hand side of the Einstein's equation becomes $\frac{8\pi G}{c^4}\,(T_{\phi\phi}-\frac{1}{2}\,g_{\phi\phi}T) = \frac{1}{2}\,\lambda G \rho^2e^{-\mu}(3p\lambda+\sigma)\approx \frac{1}{2}\,\rho^2e^{-\mu}\lambda G\sigma$ for an ideal fluid.}, i.e. 
\begin{equation}
\square \mu = 8\pi G \lambda \sigma \,.
\label{eq:wws}
\end{equation}
So, $\mu$ is $O(\lambda)$ plus possibly a constant. (Since $[G]=\siun{m}^3\cdot\siun{s}^{-2}\cdot\siun{kg}^{-1}$, $[\sigma]=\siun{kg}\cdot\siun{m}^{-3}$, $[\lambda]=\siun{m}^{-2}\cdot\siun{s}^2$, $\mu$ is indeed dimensionless.) However, by adding constants to $\mu$ and $\nu$ in exact solutions \re{metric}, one introduces or removes conical singularities (``strings'' or ``struts'') along the $z$ axis (see, e.g., \CITE{Bicak-TGS}). These, in generic cases, are attached to particles and cause their motion, as indicated in Fig.~\ref{fig:1}. If they should not represent just conical singularities introduced ``ad hoc'', they turn out to be proportional to the products $mA$, the mass and the acceleration parameters of the particles (see examples below). These products are $\sim \lambda^2$ and can thus be neglected in the Newtonian limit. Hence, we can assume $\mu\sim O(\lambda)$ and $\nu\sim O(\lambda)$ without additional constants. The axis is thus regular in the Newtonian limit, except for places where the particles occur.

Before performing the limit $\lambda\rightarrow 0$ it is crucial that we make the substitution \re{subs}. Then we find that, as $\lambda\rightarrow 0$, the metrics $t_{\alpha\beta}$ and $s^{\alpha\beta}$ approach
\begin{equation}
t_{\alpha\beta} = \diag (e^{\tilde{\mu}},\,0,\,0,\,0)\,, \quad s^{\alpha\beta} = -\diag (0,\,e^{-\tilde{\nu}},\,e^{-\tilde{\nu}},\,e^{\tilde{\mu}}/\rho^2)\,,
\label{eq:metric-limit}
\end{equation}
where $\tilde{\mu} = \lim_{\lambda\rightarrow 0} \mu( \rho^2,\,( \zeta+\frac{1}{\lambda g} )^2-t^2/\lambda)$ and analogically $\tilde{\nu}$. These limits indeed give $\tilde{\mu}=\tilde{\nu}=0$, as we discussed above, but we leave metrics in the form \re{metric-limit} to show explicitly how substitution \re{subs} alters $\mu$ and $\nu$ after the limiting process.

Starting out from \re{metric-mat} and \re{invmetric-mat}, we can calculate the affine connection and so find 
\begin{equation}
\Gamma^a_{tt} = \lim_{\lambda\rightarrow 0}  \left[ \left. \left( \frac{1}{2} \frac{\mu_{,a}}{\lambda}\right)\right|_{z=\zeta+\frac{1}{\lambda g}} \right]\,, \qquad a=\{\rho,\,z\}\,,
\label{eq:CF}
\end{equation}
(there are also non-vanishing components $\Gamma^\rho_{\phi\phi}$ and $\Gamma^\phi_{\rho\phi}$ arising due to the cylindrical coordinates -- these are not needed here). Notice that at this point the metric function $\nu$ no longer enters the Newtonian results.

As a consequence of Eq.~\re{CF} we identify the Newtonian gravitational potential as 
\begin{equation}
\Phi = \lim_{\lambda\rightarrow 0} \frac{\mu\left( \rho^2,\,\left( \zeta+\frac{1}{\lambda g} \right)^2-t^2/\lambda \right)}{2\lambda}\,.
\label{eq:np}
\end{equation}
(No ``minus'' sign arises here because of our signature.) 
This is consistent with the affine connection \re{Aff} in the sense that we obtain \re{CF} from \re{Aff}, with $t$ and $s$ given by \re{metric-limit}.

In flat spacetime with Minkowski metric $(1/\lambda,\,-1,\,-1,\,-1)$, two symmetrically located point particles move along the worldlines $\bar{x}^\mu_e=(\bar{t}(\tau),\,0,\,0,\,\bar{z}_e(\tau))$ with 
\begin{equation}
\bar{z}_e(\tau) = e \, \frac{1}{\lambda g}\,\cosh \left( \sqrt{\lambda}g\tau \right)\,,\qquad \bar{t}(\tau) = \frac{1}{\sqrt{\lambda} g}\,\sinh \left( \sqrt{\lambda}g\tau \right)\,,
\label{eq:pt}
\end{equation}
where $e=\pm 1$ denotes the particular hyperbola. Their 4-velocities are normalized to $1/\lambda$ with respect to Minkowski metric. The matter densities are thus given by $\sigma_e=\int\delta^{(4)}(x^\mu-\bar{x}_e^\mu) \,\d \tau$. After a short calculation we obtain the total density
$$\sigma = \sigma_{+1}+\sigma_{-1} =  \frac{2}{\lambda g}\, \delta\left( z^2-\frac{t^2}{\lambda}-\frac{1}{\lambda^2 g^2} \right)\,,$$
which, after substitution \re{subs} and limit $\lambda \rightarrow 0$, yields $\sigma=\delta( \zeta -\frac{1}{2}gt^2)$, i.e., we arrive at the expected Newtonian result for the source given by one uniformly accelerated particle.

Before turning to the Newtonian limit of the field, we should prove that this form of motion is not merely a peculiar coordinate artefact. We shall show that the motion takes place with respect to privileged (``rigid'') observers. We started with the ``observer field'' $u^\mu=(1/\sqrt{t_{00}},\,0,\,0,\,0)$ which is normalized to $1$ with respect to the temporal metric $t_{\alpha\beta}$ (or to $1/\lambda$ with respect to $g_{\alpha\beta}$). Following the Theorem 2.2 and Corollary 2.3 of \CITE{lazkoz-2004-460}, the important information is contained in the tensor
$$F_{\iota\kappa} = \lim_{\lambda\rightarrow 0} \frac{1}{\lambda} \,u^\bullet_{[\kappa,\iota]}\,,\qquad \text{or} \qquad F = \lim_{\lambda\rightarrow 0} \frac{1}{\lambda} \, \d u^\bullet\,,$$ 
where $u^\bullet=t_{\alpha\beta}u^\alpha \d x^\beta$, the bullet indicates that indices are lowered by temporal metric $t_{\alpha\beta}$. The observer field has the limit $u^\mu=(1,\,0,\,0,\,0)$ which remains normalized with respect to the temporal metric. This represents the field measured by static (test) observers flowing only in the absolute time $t$. They perceive a gravitational force $\vec{F}=-\nabla \Phi$ caused by potential $\Phi$  which depends on $z-\frac{1}{2}gt^2$ and thus represents the potential of a moving source.

We shall now demonstrate that in the limit $\lambda\rightarrow 0$ the field equation \re{wws} implies the standard Poisson equation for the potential $\Phi$. Introducing new variables $a=A$ and $b=B-\frac{1}{\lambda^2g^2} = z^2-t^2/\lambda-\frac{1}{\lambda^2g^2}$, we find Eq.~\re{wws} with a source $\sigma=\sigma(a,\,b)$ to become\footnote{The flat-space d'Alembert operator \re{wave-eq} is in terms of coordinates $(A,\,B)$ given by $\square = 4\left( \p_{,A}+A\p_{,AA}+\p_{,B}+B\p_{,BB} \right)$.}  
\begin{equation}
4\left[a\mu_{,aa}+\mu_{,a}+\left( b+{\textstyle \frac{1}{g^2\lambda^2}} \right)\mu_{,bb}+\mu_{,b}\right]=8\pi G\lambda\sigma\,,
\label{eq:s1}
\end{equation}
where $\p_{,A}=\p_{,a}$ and $\p_{,B}=\p_{,b}$ because the variables differ just by shift $\frac{1}{g^2\lambda^2}$. Using $\p_{,b} = \lambda\p_{,\lambda b}$, and going back to $z$ and $t$, the last equation reads
\begin{multline}
 a\mu_{,aa}+\mu_{,a}+\lambda\left(z^2\lambda-t^2 \right)\frac{\p^2\mu}{\p \left( z^2\lambda-t^2-\frac{1}{\lambda g^2} \right)^2}\\+\lambda \frac{\p\mu}{\p \left( z^2\lambda-t^2-\frac{1}{\lambda g^2} \right)} =2\pi G\lambda\sigma\,.
\label{eq:s2}
\end{multline}
Taking now the substitution \re{subs} into account, we find the limiting values of the factors entering this equation as follows: $\lim_{\lambda\rightarrow 0} \lambda\left(z^2\lambda-t^2 \right)  = \frac{1}{g^2}$ and $\lim_{\lambda\rightarrow 0} (z^2\lambda-t^2 -\frac{1}{g^2\lambda} )= \frac{2}{g}\left( \zeta-\frac{1}{2}gt^2 \right)$. Dividing \re{s2} by $\lambda$, setting, $\Phi = \lim_{\lambda\rightarrow 0} \mu/2\lambda$ following \re{np}, and putting $\Sigma = \lim_{\lambda\rightarrow 0} \sigma$, we find Eq.~\re{s2} in the limit $\lambda\rightarrow 0$ to go over to 
\begin{equation}
2\left(a\Phi_{,aa} + \Phi_{,a}\right) + \frac{1}{2}\frac{\p^2 \Phi}{\p \left( \zeta-\frac{1}{2}gt^2 \right)^2} = 2\pi G\Sigma\,. 
\label{eq:s3}
\end{equation}
After introducing finally $\bar{\zeta}=\zeta-\frac{1}{2}gt^2$ and writing back $a=\rho^2$, we recover standard Poisson's equation:
$$\triangle \Phi =\Phi_{,\rho\rho}+\frac{1}{\rho}\Phi_{,\rho}+\Phi_{,\bar{\zeta}\bar{\zeta}} = 4\pi G \Sigma\,,$$ 
where $\Phi=\Phi\left( \rho,\, \bar{\zeta}\right)$ and $\Sigma=\Sigma\left( \rho,\, \bar{\zeta} \right)$.

These results are in strong contrast with those obtained in \CITE{lazkoz-2004-460}. There the Newtonian ``sources'' -- and thus also potentials -- depend only on $\rho^2,\,t^2$ but not on $z$; hence the ``sources'' are cylindrical as, in fact, admitted in \CITE{lazkoz-2004-460}.

\section{Examples} \label{sec:examples}
In this section we wish to illustrate previous results on some typical classes of the boost-rotation symmetric spacetimes. We give the Newtonian limits for two accelerated ``monopole particles'', for Bondi's chasing freely falling particles\footnote{For simplicity and as is common in literature, we are using the term \emph{``particle''} throughout this paper although for the static Curzon-Chazy particles the ``inner'' structure of the singularity is quite complicated \CITE{1986GReGr..18..557S} (this will not change by accelerating them). Thus they are not point-like particles in an ordinary sense.} with a negative and a positive mass, for accelerated ``dipole particles'', for freely falling particles in an external field, and for the C-metric.

\subsection{Accelerated ``monopole particles''}
These are described by the Bonnor-Swaminarayan solution \CITE{BSzp} given by metric \re{metric} with
\begin{eqnarray}
\mu &=& -\frac{2\lambda Gm_1}{A_1R_1} -\frac{2\lambda Gm_2}{A_2R_2}+ 4\lambda Gm_1A_1+ 4\lambda Gm_2A_2 \,, \label{eq:bss-mu}\\
\nu &=& \lambda^2 G^2\left[-\left( \frac{m_1^2}{A_1^2R_1^4}+\frac{m_2^2}{A_2^2R_2^4}\right) \rho^2\left( z^2-t^2/\lambda \right) - \frac{2m_1m_2}{A_1A_2R_1R_2} \right. \nonumber\\
&&\qquad\left.+ 8m_1m_2\,\frac{A_1^3A_2^3}{\left( A_1^2-A_2^2 \right)^2}\frac{\left( R_1-R_2 \right)^2}{R_1R_2} \right] \nonumber \\
&&\qquad\qquad+2\lambda G\left( \frac{m_1A_1}{R_1}+\frac{m_2A_2}{R_2} \right) \left( \rho^2+z^2-t^2/\lambda  \right)\,,
\label{eq:bss-nu}
\end{eqnarray}
where
\begin{equation}
R_i = \frac{1}{2}\sqrt{\left( \rho^2+z^2-t^2/\lambda-\frac{1}{A_i^2} \right)^2+\frac{4\rho^2}{A_i^2}}\,.
\label{eq:R}
\end{equation}
In functions \re{bss-mu}\,--\,\re{R} we already inserted $\lambda$ and $G$ so as they have the correct form in standard units. In addition, we have to make substitution\footnote{These $A$'s are acceleration parameters of the solution, not the variables as in equations \re{nuA}\,--\,\re{nuB}. It should be clear from the context what is meant by $A$ in a particular equation; in this section $A$ denotes only accelerations.} $A_i=\lambda g_i$, where accelerations $g_i$ have usual dimension $\siun{m}\cdot\siun{s}^{-2}$. Since now dimensions are $[R_i]=\siun{m}^2$, $[m_i]=\siun{kg}$, $[\lambda]=\siun{m}^{-2}\cdot\siun{s}^2$, $[g_i]=\siun{m}\cdot\siun{s}^{-2}$ and $[G]=\siun{m}^{3}\cdot\siun{s}^{-2}\cdot\siun{kg}^{-1}$ the function $\mu$ is dimensionless.

This metric represents, in fact, \emph{two pairs} of particles. Only two symmetrically located particles accelerated uniformly in opposite directions with acceleration $A_1$ are obtained by putting $m_2=0$.

Regarding now the general result \re{np} for the Newtonian potential and using for its evaluation \re{bss-mu} and \re{bss-nu}, we find 
$$\Phi = \lim_{\lambda\rightarrow 0} \frac{\mu}{2\lambda}= -\frac{Gm_1}{\sqrt{\rho^2+\left( \zeta-\frac{1}{2}g_1t^2 \right)^2}}\,.$$
This is precisely the potential due to a single point particle with mass $m_1$ that moves with constant acceleration $g_1$.

In the limit $\lambda\rightarrow 0$, function $\nu$ and its derivatives go to zero as they should; in contrast to $\mu$, function $\nu$ does not enter the Newtonian limit (cf. text below Eq.~\re{CF}). Recall also (cf., e.g., Fig.~\ref{fig:roof}) that our limiting procedure removes the symmetrically located (``left'') particle to $\zeta\rightarrow -\infty$.

The case of two pairs of particles is more complicated. The particles have different accelerations and the substitution \re{subs} does not stop the second particle to run away to infinity as $\lambda$ goes to zero. However, we can keep the particles of the ``right'' pair at constant mutual distance in the hyperplane $t=0$ by introducing a length scale $L$. This can be done by setting
\begin{equation}
g_2 = \frac{1}{\frac{1}{g_1}+L\lambda}\,.
\label{eq:g2}
\end{equation}
In this case, we find that function $\mu$ in Eq.~\re{bss-mu} with both $m_1$ and $m_2$ non-vanishing, substituted into Eq.~\re{np}, implies the Newtonian potential to be
\begin{equation}
\Phi = -\frac{Gm_1}{\sqrt{\rho^2+\left( \zeta-\frac{1}{2}g_1t^2 \right)^2}}-\frac{Gm_2}{\sqrt{\rho^2+\left( \zeta-\frac{1}{2}g_1t^2 -L \right)^2}}\,.
\label{eq:phi2}
\end{equation}
This represents the potential of two particles with masses $m_1$ and $m_2$ undergoing the same acceleration $g_1$, and moving on parabolae differing only by the shift $L$.

There exist an interesting situation in the case of two pairs of particles, considered first by Bondi \CITE{bondi-nm}, when in the full relativistic solution contained in Eqs.~\re{bss-mu}\,--\,\re{bss-nu} no singularities along the $z$ axis are present. This can be achieved by choosing masses to be
$$m_1 = \frac{\left( g_1^2-g_2^2 \right)^2}{G\lambda^2g_1^3g_2^2}\,, \qquad m_2 = -\frac{\left( g_1^2-g_2^2 \right)^2}{G\lambda^2g_1^2g_2^3}\,.$$
The whole sequence of spacetimes $\mathscr{M}(\lambda),\,g_{\alpha\beta}(\lambda)$ has the $z$ axis regular. In the Newtonian limit we again obtain the potential \re{phi2} but now, as a consequence of Eq.~\re{g2}, the masses are $m_1=L^2g_1/G$ and $m_2=-L^2g_1/G$, i.e., their magnitude is the same, only signs differ. The accelerations are again same. This is the \emph{exact solution of the Newton's equations of motion} for a pair of particles with opposite masses which accelerate themselves uniformly in a straight line.

\subsection{Freely falling particle in an external field}
\label{sec:acc}
Rather surprisingly, by a similar method we are able to find the meaningful Newtonian limit even in the case of generalized Bonnor-Swaminarayan solutions which are not asymptotically flat. These solutions were constructed in \CITE{1983RSPSA.390..397B} by removing the ``outer'' particles described by the solution \re{bss-mu}\,--\,\re{bss-nu} to infinity, while increasing at the same time their mass,  so that there remain only the ``inner'' -- causally independent -- two particles, each of them falling in an external field. There are \emph{no} singularities along the $z$-axis, except for the places where these particles occur. 
\emph{Removing the ``outer'' particle as in \CITE{1983RSPSA.390..397B}, but starting from Eqs.~\re{bss-mu}\,--\,\re{bss-nu} in which constants $G$ and $\lambda$ are inserted in appropriate places, we find the generalized Bonnor-Swaminarayan solution to be given by the following functions $\mu$ and $\nu$:}
\begin{eqnarray}
\mu &=& -\frac{2mG\lambda}{AR}+A^2\left( \rho^2-z^2+t^2/\lambda +\frac{1}{A^2}\right)\,,\\
\nu &=& -\left( \frac{m^2}{A^2R^4}+\frac{A^4}{\lambda^2 G^2} \right)\rho^2\left( z^2-t^2/\lambda \right)\lambda^2 G^2 \nonumber\\
&&\qquad+A^2\left( \rho^2+z^2-t^2/\lambda -\frac{1}{A^2}\right)+\frac{2m}{AR}\left( 2A^2\rho^2+1 \right)\lambda G\,. 
\label{eq:gbs}
\end{eqnarray}
Making again substitution $A\rightarrow \lambda g$, we find that our general result \re{np} implies the Newtonian potential in the form
$$\Phi = \lim_{\lambda\rightarrow 0} \frac{\mu}{2\lambda} = -\frac{Gm}{\sqrt{\rho^2+\left( \zeta-\frac{1}{2}gt^2 \right)^2}}-g \zeta+\frac{1}{2}g^2t^2\,.$$
The potential does not now decay at infinity, however it can, indeed, be interpreted as the potential of a point particle with mass $m$ moving in a homogeneous external gravitational field of strength $g$ (purely time-dependent last term does not, of course, contribute to the field strength). Hence, a solution of Einstein's field equations representing a moving object in an external field leads, in the Newtonian limit, to a simple ``intuitive'' solution of Newton's theory.

\subsection{Self-accelerating ``dipole particles''}
This is the solution obtained from Bonnor-Swaminarayan solutions by considering two particles (with masses of opposite sign) of each pair with conical singularities located only between the particles. A limiting procedure then brings the particles in each pair together and, simultaneously, the magnitudes of the masses are increased \CITE{1983RSPSA.390..411B}. (A similar procedure produces standard dipoles in electromagnetism.) The resulting spacetime contains no conical singularities, just two causally independent self-accelerating dipole particles. The metric is given by \re{metric}, in which
\begin{eqnarray}
\mu &=& -\frac{\mathscr{D}\lambda G}{R} + \frac{\mathscr{D}\lambda G}{4A^2R^3}\left( \rho^2-z^2+t^2/\lambda +\frac{1}{A^2} \right),\\
\nu &=& -\frac{\mathscr{D}^2\lambda^2 G^2}{64R^8} \, \rho^2\left( z^2-t^2/\lambda  \right)\left[ \left( \left( \rho^2+z^2-t^2/\lambda  \right)^2-\frac{1}{A^4} \right)^2\right. \nonumber\\
&&\qquad\left.-\frac{2}{A^4}\,\rho^2\left( z^2-t^2/\lambda  \right) \right] \nonumber\\
&& \qquad\qquad-\frac{\mathscr{D}\lambda G}{4R^3}\left( \rho^2+z^2-t^2/\lambda  \right)\left( \rho^2-z^2+t^2/\lambda +\frac{1}{A^2} \right)\,,
\label{eq:dip}
\end{eqnarray}
$A$ is the magnitude of the acceleration of the symmetrically located dipoles, $\mathscr{D}$ is their constant dipole moment\footnote{In the notation of the present paper, the dipole moment is defined by $\mathscr{D}=2(m_1/A_1+m_2/A_2)$. Assuring the correct position of the struts, the dipole moment reads $\mathscr{D}=2m_1\left( A_2^2-A_1^2 \right)/A_1A_2^2$. In the limit $A_2\rightarrow A_1$ the mass parameter $m_1$ has to be rescaled to keep the dipole moment $\mathscr{D}$ constant.} with dimension $[\mathscr{D}]=\siun{kg}\cdot\siun{m}$. To obtain the Newtonian limit, we again put $A\rightarrow \lambda g$ and, after using the $\lambda$-dependent shift \re{subs} of the $z$ coordinate as in Eq.~\re{np}, we obtain the resulting limit in the form
$$\Phi = \lim_{\lambda\rightarrow 0} \frac{\mu}{2\lambda} = -\frac{\mathscr{D}G\left( \zeta-\frac{1}{2}gt^2 \right)}{\left(\rho^2+\left( \zeta-\frac{1}{2}gt^2 \right)^2\right)^{3/2}}\,.$$
This is the Newtonian potential of a uniformly accelerated dipole.

\subsection{The C-metric}
Let us turn to the last and the most elaborate example -- the C-metric. The C-metric can be interpreted as two uniformly accelerated black holes with the accelerations caused by cosmic strings. It is an example of (perhaps most frequently used) boost-rotation symmetric solution (see, e.g., \CITE{1999PhRvD..60d4004B}, for a number of references on the C-metric). In the canonical boost-rotation coordinates, it is described by ``accelerated'' rods, representing the event horizons of black holes. In metric \re{metric}, functions $\mu$, $\nu$ are now given by \CITE{bonnorscm}
\begin{eqnarray}
\hspace{-2em} \mu &=&  \ln \frac{ R_1+R_2 -\left( \tfrac{1}{2A_2^2}-\tfrac{1}{2A_1^2} \right)}{R_1+R_2 +\left( \tfrac{1}{2A_2^2}-\tfrac{1}{2A_1^2} \right)}\,,\\
\hspace{-2em}\nu &=& -\mu + \ln \frac{ \left(R_1+R_2\right)^2 -\left( \tfrac{1}{2A_2^2}-\tfrac{1}{2A_1^2} \right)^2}{4R_1R_2} + \ln\left( \frac{R_2+Z_2}{R_1+Z_1} \right)\nonumber\\
\hspace{-2em}&& \hspace{-1em}+\ln \frac{\left(\rho^2+z^2-t^2/\lambda \right)R_1+2\rho^2\left(z^2-t^2/\lambda \right)+\left(z^2-t^2/\lambda -\rho^2\right)Z_1}{\left(\rho^2+z^2-t^2/\lambda \right)R_2+2\rho^2\left(z^2-t^2/\lambda \right)+\left(z^2-t^2/\lambda -\rho^2\right)Z_2}\,,
\label{eq:CmetricBRS}
\end{eqnarray}
where $R_i$ are defined by \re{R} and $Z_i$ ($i=1,\,2$) read
$$Z_{i} = \frac{1}{2}\left( -\rho^2+z^2-t^2/\lambda -\frac{1}{A_{i}^2} \right)\,.$$
The rods in the hyperplane $t=0$ are segments on the $z$-axis given by intervals $(1/A_1,\,1/A_2)$ and $(-1/A_1,\,-1/A_2)$, $A_1>A_2$.

Since the Newtonian limit of a Schwarzschild black hole is a point mass, we anticipate the Newtonian limit of the C-metric to be an accelerated point particle. The rod should thus shrink to a point. This is achieved by putting
\begin{equation}
A_{1,2} = \frac{1}{\sqrt{\frac{1}{g^2\lambda^2}\mp\frac{2mG}{g}}}\,.
\label{eq:cmA}
\end{equation}
This is more complicated substitution for the acceleration parameter than in the previous cases. However, notice that for small mass parameter $m$ we again get $A\sim g\lambda$ as before. The parameter $m$ enters the expression for $A_{1,2}$ since it is related directly to the ``length'' of the accelerated rods. Recall that in the static case in Weyl's coordinates the Schwarzschild black hole is represented by a rod of the length $2m$ where $m$ is the Schwarzschild mass.

After long but straightforward calculations of the limit \re{np}, we obtain the resulting Newtonian potential. Remarkably, complicated form of the relativistic C-metric leads to the simple classical potential of uniformly accelerated point particle with mass $m$:
$$\Phi = -\frac{Gm}{\sqrt{\rho^2+\left( \zeta-\frac{1}{2}gt^2 \right)^2}}\,.$$

\section{Conclusions}\label{sec:conclusions}
We have constructed physically plausible Newtonian limit of general asymptotically flat boost-rotation symmetric spacetimes. We have also seen that a satisfactory Newtonian limit can be obtained in some cases which are not asymptotically flat. 

In case of relativistic spacetimes without conical singularities, when particles fall freely under mutual gravitational interaction or in an external field, the Newtonian limit produces exact self-consistent solutions of the field (Poisson) equation and of the equations of motion in Newton's theory. This satisfying result increases significantly the physical meaning of the boost-rotation symmetric spacetimes.

There exist other solutions for which we can apply the above method of finding the Newtonian limit and so extend our results. For example, one can explicitly write down functions $\mu$ and $\nu$ representing two (or even more) pairs of Schwarz\-schild black holes, Curzon particles, etc. \CITE{lunst}. Also, the solution representing a Schwarzschild black hole moving in an external field is available -- the ``generalized'' C-metric \CITE{1976JMP....17..515E}, which is the analogue of the generalized Bonnor-Swaminarayan solution discussed in Section~\ref{sec:acc}. Their Newtonian limits can be expected to be of the same form as those obtained above.

In our procedure of performing the Newtonian limit, we lost the track of the three quadrants of the original relativistic spacetime, namely, $c^2t^2>z^2$ and $c^2t^2<z^2$ for $z<0$; hence, the particle(s) accelerated in opposite direction along the negative part of $z$-axis ``disappeared'' in the limit. In Appendix, we indicate that an analogous result arises in the Newtonian limit of the complete Schwarzschild-Kruskal spacetime. In fact, the region ``preserved'' during our limiting procedure is that in which, in the original spacetime, the boost Killing vector field is timelike and the moving sources occur. From the Newtonian perspective, this region is far more intuitive than the ``radiative'' region above the roof considered as ``the natural arena'' for discussion of the Newtonian limit in \CITE{lazkoz-2004-460} because ``the region below the roof gets squeezed by the roof''.

The existence of physically ``correct'' Newtonian limits of the boost-rotation symmetric spacetimes, in particular of those in which only particles/black holes occur without any strings or struts, makes it plausible to assume that the existence of complete global relativistic solutions can be proven, at least when the fields will be ``close'' to their Newtonian counterparts. Sources will be represented by non-singular extended bodies. Bondi \cite{bondi-nm} discussed spatially extended material ``balls'' of a positive and negative mass uniformly accelerated by mutual gravity in the region below the roof, but did not succeed in constructing exact solutions. It is conceivable that when elastic bodies are considered, the existence of near-Newtonian relativistic solutions can be proven \CITE{schmidt-pc}.

In this journal, Antoci \emph{et al.} \cite{Antoci} recently questioned the physical meaning of the boost-rotation symmetric spacetimes, arguing that the acceleration horizons in these solutions (the ``roof'' in our terminology) are singular. In their Conclusion, they also found ``problematic'' the common interpretation that the singularities in these solutions represent the masses exhibiting a uniformly accelerated motion. Their claim concerning the acceleration horizons was shown to be unfounded in a detailed analysis of MacCallum \cite{2006GReGr..38.1887M}. The interpretation of singularities as accelerated sources can be supported in various ways (see, e.g., \CITE{Bicak-TGS}). An important argument for this interpretation comes from the Newtonian limit considered above.

We are grateful to the Albert Einstein Institute, Golm, for the kind hospitality, and the Grants No LC06014 (``Center of Theoretical Astrophysics'') and MSM0021620860 of the Ministry of Education of the Czech Republic for the partial support. J.B. also acknowledges the support of the Alexander von Humboldt Foundation and from the Grant GA\v{C}R 202/06/0041. Last but not least, J.B. thanks J\"{u}rgen Ehlers and Berndt Schmidt for useful discussions.

\appendix
\section{Appendix: The Newtonian limit of the Schwarzschild-Kruskal solution}\label{sec:app}
We sketch the Newtonian limit of the Schwarzschild solution in the global Kruskal coordinates. Our Newtonian limit of the boost-rotation symmetric solutions and the character of Killing horizons show some similarities. The Schwarzschild metric in standard coordinates is well suited for the Newtonian limit -- see \CITE{0264-9381-14-1A-010}:
\begin{equation}
\d s^2 = -\frac{1}{\lambda}\left( 1-\frac{2m\lambda G}{r} \right) \d t^2 + \frac{1}{1-\frac{2m\lambda G}{r}}\: \d r^2  + r^2\,\d \Omega^2 \,.
\label{eq:Schw1}
\end{equation}
Let us remind the standard procedure of introducing global Kruskal coordinates \CITE{MTW}, keeping $G$ and $\lambda=c^{-2}$. Introduce tortoise coordinate $r^*$ by
\begin{equation}
r^* = r+2m\lambda G\,\ln\left( \frac{r}{2m\lambda G}-1 \right)\,,
\label{eq:Schw-tort}
\end{equation}
and then the null coordinates $\tilde{U}$ and $\tilde{V}$ by $\tilde{V}-\tilde{U} = 2r^*$ and $\tilde{V}+\tilde{U} = 2t/\sqrt{\lambda}$. Next, coordinates $\tilde{u}$ and $\tilde{v}$ are given by $\tilde{u} = \exp(-\frac{\tilde{U}}{4m\lambda G})$ and $\tilde{v} = \exp(-\frac{\tilde{V}}{4m\lambda G})$. Finally,  introduce global time $T$ and coordinate $R$ by $\tilde{v}-\tilde{u} = \frac{2R}{m\lambda G}$ and $\tilde{v}+\tilde{u} = \frac{2T}{m\lambda\sqrt{\lambda}G}$. Then
\begin{eqnarray}
R &=& m\lambda G \, \sqrt{\frac{r}{2m\lambda G}-1}\ e^{\frac{r}{4m\lambda G}} \cosh\left(\frac{t}{m\lambda^{3/2} G}\right) \,, \label{eq:Schw-RR}\\
T &=& m\lambda^{3/2} G\, \sqrt{\frac{r}{2m\lambda G}-1}\ e^{\frac{r}{4m\lambda G}} \sinh \left(\frac{t}{m\lambda^{3/2} G}\right)\,.
\label{eq:Schw-RT}
\end{eqnarray}
These coordinates have correct dimensions of length and time.

The Schwarzschild metric \re{Schw1} becomes 
\begin{equation}
\d s^2 = \frac{32 m\lambda G}{r} e^{-\frac{r}{2m\lambda G}} \left( -\frac{1}{\lambda} \d T^2 + \d R^2 \right)+r^2\d\Omega^2\,. 
\label{eq:Schw-m}
\end{equation}
It is invariant under Lorentz transformation mixing coordinates $(R,\,T)$ together since $R^2-T^2/\lambda = m^2\lambda^2G^2\left[ r/2m\lambda G-1 \right]\exp\left( r/2m\lambda G \right)$ is unchanged.

The inverse of relations \re{Schw-RR}\,--\,\re{Schw-RT} cannot be expressed explicitly, so $r$ is to be understood as the implicit function of $R$ and $T$. It is useful to express the inversion in terms of Lambert $\lambert$ function\footnote{Lambert $\lambert$ function is the solution of the equation $x=\lambert(x) \exp\left( \lambert(x) \right)$. For details about Lambert $\lambert$ function see \CITE{corless93lamberts}.} as
\begin{eqnarray}
r &=& 2m\lambda G \left[ \lambert\left( \frac{R^2-T^2/\lambda}{m^2\lambda^2G^2e} \right)+1 \right]\,,\label{eq:rinv}\\
t &=& m\lambda^{3/2}G\arcsinh\left( \sqrt{\frac{R^2}{R^2-T^2/\lambda}-1}\ \right)\,.\label{eq:tinv}
\end{eqnarray}
The Schwarzschild metric \re{Schw-m} takes the form ($w$ denotes the parameter of Lambert $\lambert$ function, i.e., $w=\left( R^2-T^2/\lambda \right)/ m^2\lambda^2G^2e=-\tilde{u}\tilde{v}/e$):
\begin{equation}
\d s^2 = \frac{16}{e}\frac{\lambert (w)}{\lambert (w) +1} \frac{1}{w}\left( -\frac{1}{\lambda}\,\d T^2+\d R^2 \right)+ 4m^2\lambda^2G^2\left[ \lambert (w) +1\right]^2 \d \Omega^2 \,.
\label{eq:Schw2}
\end{equation}
This is wholly expressed in terms of Kruskal coordinates. For all possible values of $R$ and $T$, the value of $w$ lies in the interval $[ -1/e,\,\infty)$ where the Lambert $\lambert$ function is well-defined.  

Using the relation $\frac{\lambert (w)}{\lambert (w) +1} \frac{1}{w} = \frac{\p}{\p w} \lambert(w)$, and $w=-\tilde{u}\tilde{v}/e$, we can write
\begin{equation}
\d s^2 = \frac{16}{e}\left[ \frac{\p}{\p w} \lambert ( w )\right]\left( -\frac{1}{\lambda}\,\d T^2+\d R^2 \right)+4m^2\lambda^2G^2\left[ \lambert ( w ) +1\right]^2 \d\Omega^2 \,.
\label{eq:Schw3}
\end{equation}

\begin{figure}[!h]
\subfloat[$\lambda=1$]{
%
%
\begin{psfrags}%
\psfragscanon%
%
\psfrag{s05}[l][l]{\color[rgb]{0,0,0}\setlength{\tabcolsep}{0pt}\begin{tabular}{l}$T$\end{tabular}}%
\psfrag{s06}[l][l]{\color[rgb]{0,0,0}\setlength{\tabcolsep}{0pt}\begin{tabular}{l}$R$\end{tabular}}%
%
\psfrag{x01}[t][t]{-2}%
\psfrag{x02}[t][t]{0}%
\psfrag{x03}[t][t]{2}%
%
\psfrag{v01}[r][r]{-1}%
\psfrag{v02}[r][r]{0}%
\psfrag{v03}[r][r]{1}%
%
\includegraphics[keepaspectratio,width=2.6cm]{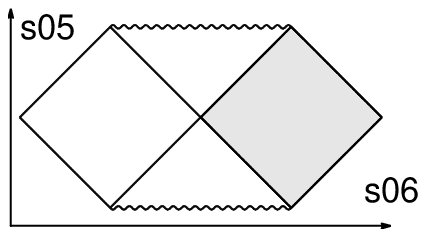}%
\end{psfrags}%
%
}\hfill
\subfloat[$\lambda=1/2$]{
%
%
\begin{psfrags}%
\psfragscanon%
%
\psfrag{s05}[l][l]{\color[rgb]{0,0,0}\setlength{\tabcolsep}{0pt}\begin{tabular}{l}$T$\end{tabular}}%
\psfrag{s06}[l][l]{\color[rgb]{0,0,0}\setlength{\tabcolsep}{0pt}\begin{tabular}{l}$R$\end{tabular}}%
%
\psfrag{x01}[t][t]{-2}%
\psfrag{x02}[t][t]{0}%
\psfrag{x03}[t][t]{2}%
%
\psfrag{v01}[r][r]{-1}%
\psfrag{v02}[r][r]{0}%
\psfrag{v03}[r][r]{1}%
%
\includegraphics[keepaspectratio,width=2.6cm]{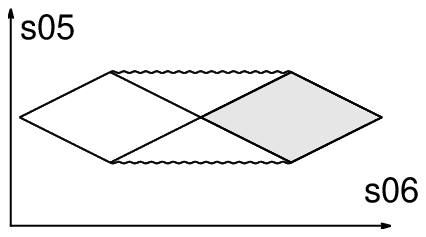}%
\end{psfrags}%
%
}\hfill
\subfloat[$\lambda=1/4$]{
%
%
\begin{psfrags}%
\psfragscanon%
%
\psfrag{s05}[l][l]{\color[rgb]{0,0,0}\setlength{\tabcolsep}{0pt}\begin{tabular}{l}$T$\end{tabular}}%
\psfrag{s06}[l][l]{\color[rgb]{0,0,0}\setlength{\tabcolsep}{0pt}\begin{tabular}{l}$R$\end{tabular}}%
%
\psfrag{x01}[t][t]{-2}%
\psfrag{x02}[t][t]{0}%
\psfrag{x03}[t][t]{2}%
%
\psfrag{v01}[r][r]{-1}%
\psfrag{v02}[r][r]{0}%
\psfrag{v03}[r][r]{1}%
%
\includegraphics[keepaspectratio,width=2.6cm]{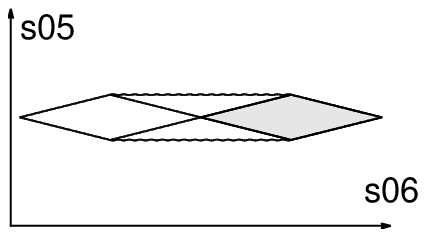}%
\end{psfrags}%
%
}\hfill
\subfloat[$\lambda=1/20$]{
%
%
\begin{psfrags}%
\psfragscanon%
%
\psfrag{s05}[l][l]{\color[rgb]{0,0,0}\setlength{\tabcolsep}{0pt}\begin{tabular}{l}$T$\end{tabular}}%
\psfrag{s06}[l][l]{\color[rgb]{0,0,0}\setlength{\tabcolsep}{0pt}\begin{tabular}{l}$R$\end{tabular}}%
%
\psfrag{x01}[t][t]{-2}%
\psfrag{x02}[t][t]{0}%
\psfrag{x03}[t][t]{2}%
%
\psfrag{v01}[r][r]{-1}%
\psfrag{v02}[r][r]{0}%
\psfrag{v03}[r][r]{1}%
%
\includegraphics[keepaspectratio,width=2.6cm]{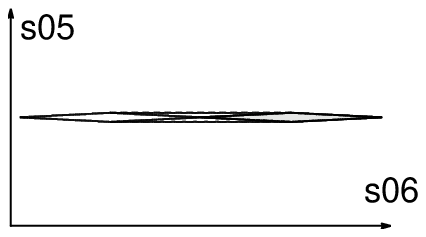}%
\end{psfrags}%
%
}
\\
\subfloat[$\lambda=1$]{
%
%
\begin{psfrags}%
\psfragscanon%
%
\psfrag{s05}[l][l]{\color[rgb]{0,0,0}\setlength{\tabcolsep}{0pt}\begin{tabular}{l}$r$\end{tabular}}%
\psfrag{s06}[l][l]{\color[rgb]{0,0,0}\setlength{\tabcolsep}{0pt}\begin{tabular}{l}$t$\end{tabular}}%
%
\psfrag{x01}[t][t]{0}%
\psfrag{x02}[t][t]{1}%
\psfrag{x03}[t][t]{2}%
%
\psfrag{v01}[r][r]{-1}%
\psfrag{v02}[r][r]{-0.5}%
\psfrag{v03}[r][r]{0}%
\psfrag{v04}[r][r]{0.5}%
\psfrag{v05}[r][r]{1}%
%
\includegraphics[keepaspectratio,width=2.6cm]{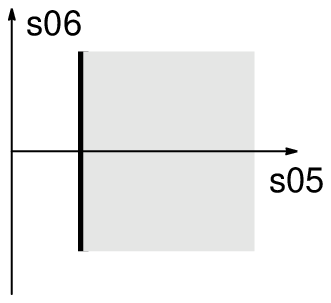}%
\end{psfrags}%
%
}\hfill
\subfloat[$\lambda=1/2$]{
%
%
\begin{psfrags}%
\psfragscanon%
%
\psfrag{s05}[l][l]{\color[rgb]{0,0,0}\setlength{\tabcolsep}{0pt}\begin{tabular}{l}$r$\end{tabular}}%
\psfrag{s06}[l][l]{\color[rgb]{0,0,0}\setlength{\tabcolsep}{0pt}\begin{tabular}{l}$t$\end{tabular}}%
%
\psfrag{x01}[t][t]{0}%
\psfrag{x02}[t][t]{1}%
\psfrag{x03}[t][t]{2}%
%
\psfrag{v01}[r][r]{-1}%
\psfrag{v02}[r][r]{-0.5}%
\psfrag{v03}[r][r]{0}%
\psfrag{v04}[r][r]{0.5}%
\psfrag{v05}[r][r]{1}%
%
\includegraphics[keepaspectratio,width=2.6cm]{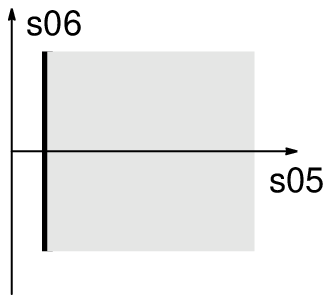}%
\end{psfrags}%
%
}\hfill
\subfloat[$\lambda=1/4$]{
%
%
\begin{psfrags}%
\psfragscanon%
%
\psfrag{s05}[l][l]{\color[rgb]{0,0,0}\setlength{\tabcolsep}{0pt}\begin{tabular}{l}$r$\end{tabular}}%
\psfrag{s06}[l][l]{\color[rgb]{0,0,0}\setlength{\tabcolsep}{0pt}\begin{tabular}{l}$t$\end{tabular}}%
%
\psfrag{x01}[t][t]{0}%
\psfrag{x02}[t][t]{1}%
\psfrag{x03}[t][t]{2}%
%
\psfrag{v01}[r][r]{-1}%
\psfrag{v02}[r][r]{-0.5}%
\psfrag{v03}[r][r]{0}%
\psfrag{v04}[r][r]{0.5}%
\psfrag{v05}[r][r]{1}%
%
\includegraphics[keepaspectratio,width=2.6cm]{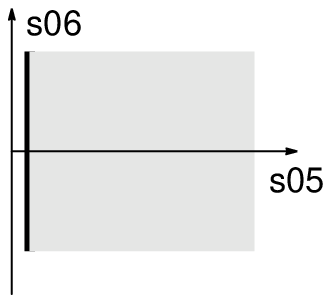}%
\end{psfrags}%
%
}\hfill
\subfloat[$\lambda=1/20$]{
%
%
\begin{psfrags}%
\psfragscanon%
%
\psfrag{s05}[l][l]{\color[rgb]{0,0,0}\setlength{\tabcolsep}{0pt}\begin{tabular}{l}$r$\end{tabular}}%
\psfrag{s06}[l][l]{\color[rgb]{0,0,0}\setlength{\tabcolsep}{0pt}\begin{tabular}{l}$t$\end{tabular}}%
%
\psfrag{x01}[t][t]{0}%
\psfrag{x02}[t][t]{1}%
\psfrag{x03}[t][t]{2}%
%
\psfrag{v01}[r][r]{-1}%
\psfrag{v02}[r][r]{-0.5}%
\psfrag{v03}[r][r]{0}%
\psfrag{v04}[r][r]{0.5}%
\psfrag{v05}[r][r]{1}%
%
\includegraphics[keepaspectratio,width=2.6cm]{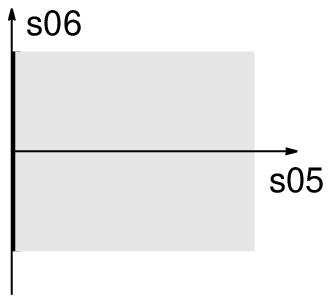}%
\end{psfrags}%
%
}
\caption{The Newtonian limit of a Schwarzschild black hole in the Kruskal coordinates $(T,R)$ (figures (a)\,--\,(d)) and in the original -- from the Newtonian point of view more physical -- Schwarzschild coordinates $(t,r)$ (figures (e)\,--\,(h)). This should be compared with Fig. \ref{fig:roof} in the main text.} 
\label{fig:schw}
\end{figure}

The analogue of the roof gets ``squeezed'' in the limit $\lambda\rightarrow 0$, whereas worldlines $r=$~const. (i.e., hyperbolas) in the Kruskal coordinates $(T,\,R)$ run away to infinity as can be seen from Eqs.~\re{Schw-RR}\,--\,\re{Schw-RT}. The worldlines of constant $r$ are worldlines of uniformly accelerated observers whose acceleration is set precisely to balance the attractive force of the central black hole; these worldlines are tangent to the Killing vector $\p/\p t$.

The structure of the Killing vector $\p/\p t$ in the Kruskal coordinates in the vicinity of the event horizon resembles the boost vector in the vicinity of acceleration horizon: $\xi^\mu \sim ( R\sqrt{\lambda},\,T/\sqrt{\lambda} )$ (although topologies differ).


\begin{thebibliography}{10}
\providecommand{\url}[1]{{#1}}
\providecommand{\urlprefix}{URL }
\expandafter\ifx\csname urlstyle\endcsname\relax
  \providecommand{\doi}[1]{DOI~\discretionary{}{}{}#1}\else
  \providecommand{\doi}{DOI~\discretionary{}{}{}\begingroup
  \urlstyle{rm}\Url}\fi

\bibitem{AGS}
Alcubierre, M., Gundlach, C., Siebel, F.: Integration of geodesics as a test
  bed for comparing exact and numerically generated spacetimes.
\newblock In: Abstracts of Plenary Lectures and Contributed Papers (GR15)
  (1997)

\bibitem{Antoci}
Antoci, S., Liebscher, D., Mihich, L.: {The physical meaning of the
  ``boost-rotation symmetric'' solutions within the general interpretation of
  Einstein's theory of gravitation}.
\newblock Gen. Relativ. Gravit. \textbf{38}, 15--22 (2006)

\bibitem{1999PhRvD..60d4004B}
{Bi{\v c}{\'a}k}, J., {Pravda}, V.: {Spinning C metric: Radiative spacetime
  with accelerating, rotating black holes}.
\newblock Phys. Rev.~D \textbf{60}, 044004, 10 pages (1999)

\bibitem{Bicak68}
Bi\v{c}{\'{a}}k, J.: {Gravitational {R}adiation from {U}niformly {A}ccelerated
  {P}articles in {G}eneral {R}elativity}.
\newblock Proc. Roy. Soc. Lond.~A \textbf{302}, 201--224 (1968)

\bibitem{bicak-ssefe}
Bi\v{c}{\'{a}}k, J.: {Selected Solutions of Einstein's Field Equations: Their
  Role in General Relativity and Astrophysics}.
\newblock In: B.G. {Schmidt} (ed.) Einstein's Field Equations and Their
  Physical Implications, \emph{Lecture Notes in Physics, Berlin Springer
  Verlag}, vol. 540, pp. 1--126 (2000)

\bibitem{1983RSPSA.390..397B}
Bi\v{c}{\'{a}}k, J., {Hoenselaers}, C., Schmidt, B.: {The solutions of the
  Einstein equations for uniformly accelerated particles without nodal
  singularities. I. Freely falling particles in external fields}.
\newblock Proc. Roy. Soc. Lond.~A \textbf{390}, 397--409 (1983)

\bibitem{1983RSPSA.390..411B}
Bi\v{c}{\'{a}}k, J., {Hoenselaers}, C., Schmidt, B.: {The solutions of the
  Einstein equations for uniformly accelerated particles without nodal
  singularities. II. Self-accelerating particles}.
\newblock Proc. Roy. Soc. Lond.~A \textbf{390}, 411--419 (1983)

\bibitem{nevim3}
Bi\v{c}{\'{a}}k, J., Pravdov{\'{a}}, A.: Symmetries of asymptotically flat
  electrovacuum space-times and radiation.
\newblock J. Math. Phys. \textbf{39}, 6011--6039 (1998)

\bibitem{BRW}
Bi\v{c}{\'{a}}k, J., Reilly, P., Winicour, J.: {Boost-rotation symmetric
  gravitational null cone data}.
\newblock Gen. Relativ. Gravit. \textbf{20}, 171--181 (1988)

\bibitem{1984JMP....25..600B}
Bi\v{c}{\'{a}}k, J., Schmidt, B.: {Isometries compatible with gravitational
  radiation}.
\newblock J. Math. Phys. \textbf{25}, 600--606 (1984)

\bibitem{Bicak-TGS}
Bi\v{c}{\'{a}}k, J., Schmidt, B.: Asymptotically flat radiative space-times
  with boost-rotation symmetry: The~general structure.
\newblock Phys. Rev.~D \textbf{40}, 1827--1853 (1989)

\bibitem{bondi-nm}
Bondi, H.: {{N}egative {M}ass in {G}eneral {R}elativity}.
\newblock Rev. Mod. Phys \textbf{29}, 423--428 (1957)

\bibitem{bonnorscm}
Bonnor, W.: {The {S}ources of the {V}acuum C-metric}.
\newblock Gen. Relativ. Gravit. \textbf{15}, 535--551 (1983)

\bibitem{BSzp}
Bonnor, W., Swaminarayan, N.: An {E}xact {S}olution for {U}niformly
  {A}ccelerated {P}articles in {G}eneral {R}elativity.
\newblock Z. Phys. A \textbf{177}, 240--256 (1964)

\bibitem{Cartan1}
Cartan, E.: {Les Vari\`{e}t\`{e}s A Connexion Affine et La Th\'{e}orie De La
  Relativit\'{e} G\'{e}n\'{e}ralis\'{e}e}.
\newblock Ann. Ecole Norm. \textbf{40}, 326--412 (1922)

\bibitem{Cartan2}
Cartan, E.: {Les Vari\`{e}t\`{e}s A Connexion Affine et La Th\'{e}orie De La
  Relativit\'{e} G\'{e}n\'{e}ralis\'{e}e}.
\newblock Ann. Ecole Norm. \textbf{41}, 1--25 (1924)

\bibitem{corless93lamberts}
Corless, R., Gonnet, G., Hare, D., Jeffrey, D., Knuth, D.: On the {L}ambert's
  {W} function.
\newblock Advances in Computational Mathematics \textbf{5}, 329--359 (1996)

\bibitem{2003CQGra..20..127D}
{Dowker}, H., {Thambyahpillai}, S.: {Many accelerating black holes}.
\newblock Class. and Quantum Grav. \textbf{20}, 127--135 (2003)

\bibitem{0264-9381-23-2-005}
Dutta, K., Ray, S., Traschen, J.: Boost mass and the mechanics of accelerated
  black holes.
\newblock Class. and Quantum Grav. \textbf{23}, 335--352 (2006)

\bibitem{0264-9381-14-1A-010}
Ehlers, J.: {Examples of Newtonian limits of relativistic spacetimes}.
\newblock Class. and Quantum Grav. \textbf{14}, A119--A126 (1997)

\bibitem{E2}
Ehlers, J.: {The Newtonian Limit of General Relativity}.
\newblock In: Understanding Physics. Coppernicus Gesellschaft e.V.,
  Katlenburg-Lindau (1998)

\bibitem{EhlersN}
Ehlers, J.: Newtonian {L}imit of {G}eneral {R}elativity.
\newblock In: Encyclopedia of {M}athematical {P}hysics, vol.~3, pp. 503--509.
  Elsevier (2006)

\bibitem{1976JMP....17..515E}
{Ernst}, F.: {Removal of the nodal singularity of the C-metric}.
\newblock J. Math. Phys. \textbf{17}, 515--516 (1976)

\bibitem{Fr}
Friedrichs, K.: {Eine invariante Formulierung des Newtonschen
  Gravitationsgesetzes und des Grenz{\"{u}}bergangs vom Einsteinschen zum
  Newtonschen Gesetz}.
\newblock Math. Ann. \textbf{98}, 566--575 (1927)

\bibitem{GPW}
G\'{o}mez, R., Papadopoulos, P., Winicour, J.: {Null cone evolution of
  axisymmetric vacuum space-times}.
\newblock J. Math. Phys. \textbf{35}, 4184--4204 (1994)

\bibitem{nevim2}
Israel, W., Khan, K.: {Collinear particles and Bondi dipoles in general
  relativity}.
\newblock Nuov. Cim. \textbf{33}, 331 (1964)

\bibitem{PhysRevD.2.1359}
Kinnersley, W., Walker, M.: {Uniformly {A}ccelerating {C}harged {M}ass in
  {G}eneral {R}elativity}.
\newblock Phys. Rev.~D \textbf{2}, 1359--1370 (1970)

\bibitem{lazkoz-2004-460}
Lazkoz, R., Valiente~Kroon, J.: The {N}ewtonian limit of spacetimes describing
  uniformly accelerated particles.
\newblock Proc. Roy. Soc. Lond.~A \textbf{460}, 995--1016 (2004)\\
See also Lazkos, R. and Valinete Kroon, J.: {The Newtonian limit of the spacetimes describing uniformly accelerated particles}. In: Proceedings of the 10th Marcel Grossmann Meeting, Part C, World Scientific, New Jersey-London-Singapore (2005)

\bibitem{2006GReGr..38.1887M}
{MacCallum}, M.: {On singularities, horizons, invariants, and the results of
  Antoci, Liebscher and Mihich (Gen Relativ Gravit 38, 15 (2006) and earlier)}
  \textbf{38}, 1887--1899 (2006)

\bibitem{MTW}
{Misner}, C., {Thorne}, K., {Wheeler}, J.: {Gravitation}.
\newblock San Francisco: W.H.~Freeman and Co. (1973)

\bibitem{2000CzJPh..50..333P}
Pravda, V., Pravdov\'{a}, A.: {Boost-rotation symmetric spacetimes - review.}
\newblock Czech. J. Phys. \textbf{50}, 333--375 (2000)

\bibitem{Rindler}
Rindler, W.: {Relativity, Special, General, and Cosmological}, {Second} edn.
\newblock {Oxford Univerisity Press}, {Oxford} (2006)

\bibitem{schmidt-pc}
Schmidt, B.: Private communication

\bibitem{1986GReGr..18..557S}
{Scott}, S., {Szekeres}, P.: \\
1. {The Curzon singularity. I: Spatial sections.} \textbf{18}, 557--570 (1986)\\
2. {The Curzon singularity. II: Global picture.} \textbf{18}, 571--583 (1986)

\bibitem{lunst}
Steele, J., Lun, A.: {{O}n {R}ational {R}epresentation of {S}tationary
  {A}xi\-symmetric {V}acuum metrics~{I}: {C}ubic and {Q}uartic Coordinates}.
\newblock Relativity Today: Proceedings of the Fourth Hungarian Workshop in
  Relativity  (1994)

\bibitem{SC}
Stephani, H., Kramer, D., MacCallum, M., Hoenselaers, C., E., H.: {E}xact
  {S}olutions of {E}instein's {F}ield {E}quations, {S}econd {E}dition.
\newblock Cambridge University Press, Cambridge (2003)

\bibitem{Trautmann}
Trautman, A.: Comparison of {N}ewtonian and {R}elativistic {T}heories of
  {S}pace-times.
\newblock In: Perspectives in geometry and relativity, p. 425. Indiana
  University Press (1966)

\end{thebibliography}
\end{document}